# The shadow system in coordinate space:
# A locally explicable system that violates Bell's inequality

Warren Leffler
Department of Mathematics, Mills College,
5000 McArthur Blvd.
Oakland, California  94613, USA
wleffler@mills.edu

## ABSTRACT

We explore the consequences of denying the "emptiness of paths not taken," EPNT, a premise of Bernstein, Greenberger, Horne, and Zeilinger in their paper titled, *Bell theorem without inequalities*. Carrying out the negation of EPNT leads to the concept of a "shadow stream." Streams are essentially particle implementations of the path amplitudes in Feynman's formulation of quantum mechanics, resulting in a simple and consistent extension of the standard postulates for quantum mechanics. Following up on the negation of EPNT using shadow streams, we analyze the experimental outcomes in a standard two-particle interferometer. The result is a simple counterexample to Bell's theorem in position/momentum space.

## I. INTRODUCTION

In an important paper for quantum foundations Bernstein, Greenberger, Horne, and Zeilinger (BGHZ) [1] employed a two-particle interferometer setup [2] to argue that the experimental outcomes for two entangled particles cannot be reproduced by a local, realistic model. Such a model (from the famous EPR paper [3]) forbids interaction between events in space-like separated regions, while also requiring that measurement outcomes reveal pre-existing properties of observables. In arguing against the EPR program and in favor of what they call "Bell theorem without inequalities," BGHZ find it necessary to augment the principles of EPR with the premise, EPNT.  EPNT stands for "emptiness of paths not taken." Regarding this they write,

> … one can deny EPNT and thereby imagine that *something* could travel down the empty beam, so as to provide information to the nonempty beam, when the two beams meet.  And this something could be consistent with EPR locality, if the particles (and these somethings) on opposite sides of the origin do not communicate.

On the basis of EPNT, BGHZ do indeed establish a "Bell theorem without inequalities." This is an important result—a valid and quite straightforward proof.

But what if (unlike BGHZ) we actually assume the negation of EPNT?  This is the point of departure for our argument. But we emphasize here that, apart from this starting point, our approach is completely independent— both conceptually and logically—of the BGHZ paper.

Accordingly, we assume the negation of EPNT. That is, we view the beams in the two-particle interferometer as containing these hypothetical "somethings," which we call "shadow particles"—whose properties were originally described by David Deutsch in a different context (a discussion of double-slit interference). [4] Rather strikingly the denial of EPNT, and its implementation via Deutsch's shadow particles, leads to a short, transparent, and elementary counterargument to Bell's theorem (the derivation could be understood by high school students).

But first note that, in an important sense, Bell's theorem is not directly about quantum mechanics at all. As John Bell explains here, [5]

> So the following argument will not mention particles, nor indeed fields, nor any other particular picture of what goes on at the microscopic level. Nor will it involve any use of the words "quantum mechanical system," which can have an unfortunate effect on the discussion. The difficulty is not created by any such picture or any such terminology. It is created by the predictions about the correlations in the visible outputs of certain conceivable experimental set-ups.

But then he quickly goes on to state (which is what all the fuss is about), "… we will argue that certain particular correlations, realizable according to quantum mechanics, are *locally inexplicable*. They cannot be explained, that is to say, without action at a distance." Again (in his 1964 paper, [6]) he defines locality as meaning "the result of a measurement on one system [is] unaffected by operations on a distant system with which it has interacted in the past." He then states that being nonlocal "is characteristic … of [any theory] which reproduces exactly the quantum mechanical predictions." He concludes this because an inequality he derives in his theorem is violated by quantum mechanical correlations involving entangled particles.

Our counterargument in Sec. III employs a standard two-particle interferometer setup (the kind used by BGHZ). Working from it we will show in a straightforward way that the correlations for spinless, entangled particles in Euclidean configuration space are (to use Bell's phrase) "locally explicable": that is, local causality operates throughout to govern the passage of the entangled particles from source event to detectors. The correlations do indeed violate Bell's inequalities, but they nevertheless involve no action at a distance. That is, we show in Sec. III that the hypotheses of Bell's theorem (in the 1964 paper cited above there were two) do not have the logical consequence that failure to satisfy the inequalities (as mentioned, inequalities derived in the theorem) implies "locally inexplicable" (action at a distance). In other words, we present a counterexample to Bell's theorem in position/momentum phase space.

The simple and basic idea of our proof, apart from the mathematical details (which themselves involve only standard elementary linear algebra), is that from the perspective of the "shadow interpretation" (SI) there are four paths and four particles involved when a source event sends two entangled particles through the arms of the two-particle interferometer to beamsplitters located in opposite wings of the device. In each wing, one path is taken by a tangible particle (ordinary particle) and one by its shadow counterpart (the negation of EPNT). There are thus four "path-amplitudes" associated with interference between tangible and shadow particles when they meet at the beamsplitters. Although of course not essential, it is helpful in this context to view the amplitudes as functions of rotating unit vectors in the complex plane, the rotation starting when a particle is emitted and continuing until it arrives at a destination point (Richard Feynman's "clocks" [7]). The effects when the particles arrive at a beamsplitter are purely local in nature, occurring via contiguous interaction, as the tangible and shadow particles interfere with each other when their paths come together at the beamsplitter. This interference determines of course where the tangible particles travel next when they leave the beamsplitter on their way to a detector. The interference is quantified in the usual way, involving the usual operations and amplitude terms, and leading to the usual result.

So far there is nothing new in this. It is all, apart from the hypothesized shadow particles, quite conventional and obvious. But what is now new when Deutsch's shadow particles are taken into account is that the shadow particles contain complementary information regarding the paths not taken by the tangible particles. Interestingly, this information is actually displayed in the standard amplitude equations (equations 3.1 and 3.2 in Sec. III). But from the perspective of the shadow system we now see that each amplitude term in such an equation corresponds to the path of a particular one of the four particles, tangible or shadow. Moreover, because of the necessary geometry in the design of an interferometer, there is information stored in the path amplitudes (or clocks, phases) that is common to both sides of the experiment. This common information occurs because of congruent paths built into the interferometer (such paths are required by the well-known "principle of path indistinguishability," although—as a corollary to our argument shows—what is really necessary in general is that the corresponding path amplitudes be equal). When a simple substitution (switch) of equal amplitudes for equals is now made into the original equation it becomes immediately obvious that it is precisely this equivalent complementary information that underlies the correlations in two-particle interference: The resulting equation is equivalent to the original one, but the new equation now describes the physical situation directly in terms of local interaction and local interference effects. Each separate tangible particle is thus seen to go where the information shared locally with the shadow particle directs it when the particles meet at a beamsplitter. The net outcome produces the usual correlations for two-particle interference, but now we see that there is no communication across the origin: no action at a distance, although the end result is as though there were.

Therefore, to reiterate an earlier point: The argument of Sec. III shows that the correlations of entangled particles in coordinate space are *locally explicable* (to use Bell's phrase, quoted earlier), and we therefore have a counterexample to Bell's theorem.

Of course for our purposes it is one thing to assume that shadow particles exist. It is quite another to describe their properties mathematically. In Sec. II we describe shadow particles a bit more fully, drawing on Deutsch's original qualitative description of them. We further assume that they obey the same dynamics as ordinary particles (as mentioned, what Deutsch calls "tangible particles"). We then quantify their interactions with tangible particles using two new postulates. These postulates are immediately seen to form a consistent extension of standard quantum mechanics (QM). For example, one postulate states that, given *any* two points, when there is more than one possible path that a tangible particle can take in traveling between the points, it *randomly* takes one of the paths and shadow particles take the other paths. Now in, say, a double-slit experiment with two entangled particles [8]—where there are uncountably many paths involved—this would play a central role in our approach to a counterexample to Bell's theorem; but in the two-particle interferometer setup that we analyze in Sec. III it only becomes relevant at the beamsplitters, though it nevertheless plays a vital role. The tangible particle together with its shadow counterparts comprise a "shadow stream," and the amplitude of the stream is just the sum of the individual path amplitudes. This is our version of Feynman's rule [9] that "the amplitude of an event is the sum of the amplitudes for the various alternative ways that the event can occur … . The total amplitude will be the sum of a contribution from each of the paths."  Thus this postulate is just a trivial translation of Feynman's amplitude formulation of quantum mechanics into our shadow system, SI. Another postulate (again a simple modification of one from Feynman) stipulates that distinct tangible particles generate distinct shadow streams, and the amplitude of the composite system is the product of the separate amplitudes (this also follows from the way amplitudes are computed in the tensor product of Hilbert spaces).  The predictions of SI, which only involve such amplitudes, necessarily agree completely with those of the standard theory.  From this point of view, the consistency of SI with the standard theory is obvious. Alternatively, one can appeal to a well-known fact from mathematical logic, which states that any set of sentences that hold in a model are consistent. [10]  Now the double-slit setup (or any standard interferometer setup) is a model of QM + SI (that is, all valid statements of the standard theory plus those of SI are satisfiable in the double-slit structure) and therefore the combined theory is consistent.  In any case, we have a consistent extension of the standard system, almost by definition.

The failure of Bell's derivation to rule out local causality in position/momentum phase space runs counter to dozens of research papers over the past several decades. Here is one famous recent example regarding a two-particle-interferometer experiment in Geneva [11] (famous because the measurements were separated by more than 10 km). The experimenters write in a later report, [12] "How can these spatially separated locations 'know' what happens elsewhere? Is it all predetermined at the source? Do they somehow communicate? … that it is all predetermined at the source, has been ruled out by numerous Bell tests. …  If some sort of hidden communication does exist, it must propagate faster than light … the hypothetical hidden communication must propagate at least $10^7$ times faster than the speed of light!"

Naturally, all serious experimental research on entangled particles has its own breathless counterpart in non-technical articles and books. There is even a popular book titled, *The Age of Entanglement: When Quantum Physics Was Reborn*.  [13] As for the Geneva experiments, here is what a recent article in the Wall Street Journal had to say: [14]

> Some philosophers see quantum phenomena as a sign of far greater unknown forces at work and it bolsters their view that a spiritual dimension exists. "We don't know how nature manages to produce spooky behavior," says Nicolas Gisin, a scientist at Geneva University, who led a recent experiment demonstrating action-at-a-distance. "But it's a fascinating time for physics because it can be mastered and exploited." … Because of its bizarre implications, quantum theory has been used to investigate everything from free will and the paranormal to the enigma of consciousness … . Several serious physicists have devoted their lives to the study of such ideas, including Bernard d'Espagnat. In March, the 87-year-old Frenchman won the prestigious $1.5 million Templeton Prize for years of work affirming "life's spiritual dimension." Based on quantum behavior, Dr. d'Espagnat's big idea is that science can only probe so far into what is real, and there's a "veiled reality" that will always elude us.

Actually, although we do not pursue the point very far in this paper, the shadow system affords a peek beneath d'Espagnat's "veil"—providing a deeper insight into the nature of quantum reality and elegantly eliminating a host of quantum paradoxes. But in any case, perhaps our counterexample to Bell's theorem, although it is contrary to a now deeply-entrenched conventional view about the theorem, is not so surprising after all. The reason has to do with the origin of the shadow system. Deutsch originally conceived of shadow particles in terms of Hugh Everett's many-worlds interpretation of quantum mechanics (MWI). Now Bell's theorem does not hold in MWI. As Lev Vaidman has put it, "in the framework of the MWI, Bell's argument cannot get off the ground because it requires a predetermined single outcome of a quantum experiment."[15] There is, however, an important difference here between our approach, SI, and MWI: The shadow interpretation is what might be called an "internal model," treating all particle interaction as taking place within a single world, the one we experience.

Of course in MWI Bell's theorem also fails for correlations involving spin effects. This suggests that the same would be true for a suitable extension of SI to accommodate spin in a way that embodies local realism. This may indeed be the case, with features common to both wings of the experiment again carrying complementary information to produce the correlations. But in Euclidean space the shadow construct is based ultimately on Feynman path integrals, whereas for spin the "paths" carrying complementary information (in the form of exponentiated "action-clocks" ticking away in coordinated fashion on each particle) would be in $\mathbb{R}^3 \times SO(3)$, where $SO(3)$ is the group manifold of rotations in Euclidean space. Paths constitute the unifying idea for both coordinate space and spin. In addition to shadow particles traveling in coordinate space, we would have hypothetical shadow entities (say, infinitely many spinning concentric shadow spheres or tops, their radii differing infinitesimally) "traveling" paths in $\mathbb{R}^3 \times SO(3)$ (path interference would eliminate the obvious paradoxes involving superluminal spin rates). The action would now have two degrees of freedom—the zenith and azimuth Euler angles—and the path integral would be over a multiply-connected space. [16, 17]

An intermediate case (where the path integral is also over a multiply-connected space, with paths classified by their winding number about a fixed point) is the so-called quantum particle on a ring. Here the phase space is the cylinder, $\mathbb{R} \times S^{(1)}$. In this case, when two entangled rings are separated in a source event (in a doubtless experimentally infeasible gedanken experiment), the basic property of the propagator (convolution of kernels) [18] ensures that common amplitudes lead to correlations, with shadow particles on each ring traveling and interfering along concentric circles whose radii differ infinitesimally. A similar idea extends to cover spin.

The details of the spin construct are beyond the scope of this paper, however, which is principally concerned with presenting a simple counterexample to Bell's theorem.

# II. POSTULATES FOR SHADOW STREAMS

… Feynman proclaimed [about the two-slit experiment] that each electron that makes it through to the screen actually goes through both slits. It sounds crazy; but hang on: Things get even more wild. Feynman argued that in traveling from the source to a given point on the phosphorescent screen each individual electron actually traverses *all possible trajectories simultaneously;* … It goes in a nice orderly way through the left slit. It simultaneously also goes in a nice orderly way through the right slit. It heads toward the left slit, but suddenly changes course and heads through the right. It meanders back and forth, finally passing through the left slit. It goes on a long journey to the Andromeda galaxy before turning back and passing through the left slit on its way to the screen. And on and on it goes: the electron, according to Feynman, simultaneously "sniffs" out *every* possible path connecting its starting location with its final destination. [19]

We begin by assuming the usual postulates for quantum mechanics (any such set will do) [≗]. We then present two additional postulates below (actually three, because for convenience we separate the first postulate into two parts). As stated in Sec. I, the postulates are essentially translations into our system of Feynman's amplitude formulation of QM. In his book with Albert Hibbs, Feynman defined the amplitude $K(b, a)$ (called the "kernel") for a non-relativistic particle traveling from $a$ to $b$ to be the sum over all amplitudes for each trajectory that goes from $a$ to $b$: [20]

$$K(b,a) = \sum_{\substack{\text{over all paths} \\ \text{from } a \text{ to } b}} \phi[x(t)], \text{ where } \phi[x(t)] = \text{const } e^{iS[x(t)]/\hbar}$$

As we know, the properties of this function lead to a formulation of QM (Feynman's path-integral or amplitude formulation) equivalent to those based on Schrödinger's wave equation or Heisenberg's matrix mechanics; and, as mentioned earlier, this amplitude formulation provides a general theoretical justification for the shadow system in Euclidean space. But in our central argument in Sec. III we will be dealing with just a two-particle interferometer, and therefore with only a few paths. In that context the elementary rules regarding quantum amplitudes given in Feynman's classic *Lectures on Physics* will suffice. We list his rules below for easy comparison with the new postulates: [21]

> (2.1) The probability $P$ of a quantum event is the square of the absolute value of a complex number $\phi$ (the probability amplitude): $P = \phi^* \phi = |\phi|^2$.

> (2.2) When an event can occur in alternative ways, the combined amplitude, $\phi$, is the sum of the amplitudes for each way considered separately, $\phi = \phi_1 + \phi_2$

> (2.3) When a particle goes by some particular route, the amplitude for that route is the product of the amplitude to go part way with the amplitude to go the rest of the way.

> (2.4) Given two non-interacting particles, "the amplitude that one particle will do one thing and the other one do something else is the product of the two amplitudes that the two particles would do the two things separately."

Feynman famously emphasized that these are just computational rules, and that "No one has found any machinery behind the law. No one can 'explain' any more than we have just 'explained.' No one will give you any deeper representation of the situation." Nevertheless, the interpretation presented here does indeed provide machinery behind the law (in terms of

---

[≗] Here is a representative online set from the Georgia State University physics website: http://hyperphysics.phy-astr.gsu.edu/Hbase/quantum/qm.html

the easily visualized and classical picture of a stream); and, although we do not discuss this further in this paper, it provides an elegant resolution of a host of single-particle interference paradoxes. But perhaps most important, on the basis of physical locality (contiguous interaction), it explains the otherwise mysterious-seeming correlations that, under certain circumstances, seem to take place instantaneously between independently performed and distantly separated measurements.

**Postulate 1: Shadow and tangible particles.** As mentioned in Sec. I, we will draw on Deutsch's shadow particles for our implementation of the hypothetical "somethings" that result from the negation of BGHZ's premise, EPNT: "emptiness of paths not taken." Deutsch argued that when an ordinary particle (a *tangible* particle) is emitted from a source to, say, a detecting screen in a double-slit experiment, the fringe pattern observed on the screen comes from the interaction of the tangible particle with shadow counterparts coming from parallel universes. The shadow particles interfere only with tangible particles of the same type, and therefore they can be detected only indirectly, through their effects on regular, tangible particles—that is, shadow photons interfere with regular photons, shadow electrons with regular electrons, and so on. Thus what we call the "shadow interpretation" in this paper is essentially a mathematical implementation in coordinate space of Deutsch's original qualitative description of shadow particles but without the assumption of parallel universes.

**Postulate 2: Streams and their amplitudes.** This is the key postulate: Whenever a single tangible particle travels from one point to another and has a choice of paths that it can take, it randomly takes one of the possible paths, and distinct shadow particles take the others. Thus, all possible paths that could be traveled by the tangible particle are "filled" (a one-to-one correspondence) by accompanying particles (the negation of EPNT). The collection of all such particles—the tangible particle and its counterpart shadow particles—is called the *shadow stream* associated with the tangible particle. The fundamental object in SI is this shadow stream. Streams are precisely what state vectors in Hilbert space describe: streams of tangible and shadow particles. Thus a tangible particle is not a wave, and it does not interfere with itself. Tangible and shadow particles comprise a stream, and interactions within the stream produce the observed interference effects. When there are infinitely many possible paths involved, the amplitude of a shadow stream generated by a single tangible particle is quantified by the corresponding Feynman path integral (i.e., using the kernel above; also see [22] and (2.5) below)—that is, the time-evolution propagator is given directly by a path integral. But when there are only finitely many possible paths for the tangible particle to take—as in, say, an interferometer—rule (2.2) above suffices to quantify the stream's amplitude: The amplitude is the sum of the individual "path-amplitudes." (Here we are using the term "path-amplitude" in a self-evident way.)

Throughout this paper for simplicity we only consider particles moving in one spatial dimension over time (extending the system to accommodate more degrees of freedom is straightforward). Fig. 2.1 below illustrates various possible shadow streams generated by a tangible particle and its accompanying shadow particles as they move over several infinitesimal time stages (the picture is the kind that the mathematician H. Jerome Keisler termed "an infinitesimal snapshot"). [23] Over an infinitesimal time interval a particle can be considered free (not subject to a potential), so its path is approximately linear—a ray.

The stage $t_k$ to $t_k + \varepsilon$ shows five shadow streams, each consisting of a single ray. From $t_k + \varepsilon$ to $t_k + 2\varepsilon$ there are three streams, each containing five rays. From $t_k + 2\varepsilon$ to $t_k + 3\varepsilon$ there are two shadow streams, each with three rays.

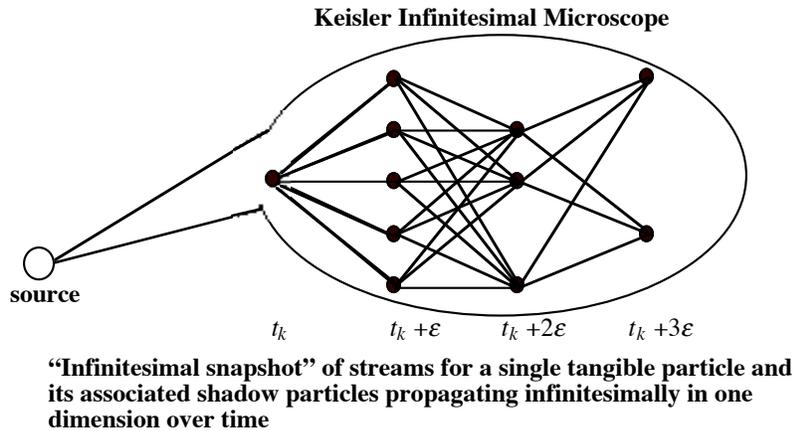

**"Infinitesimal snapshot" of streams for a single tangible particle and its associated shadow particles propagating infinitesimally in one dimension over time**

**Figure 2.1**

**Postulate 3: Streams generated by more than one tangible particle.** Here we posit that distinct tangible particles generate distinct shadow streams. To quantify the combined amplitude for the separate shadow streams we apply rule (2.4) above: multiply the separate amplitudes. Alternatively, one can take account of the fact that different streams can always be described by different Hilbert spaces in a tensor-product space. Therefore, the combined amplitude is simply the product of the separate amplitudes—which of course is the same as a sum of products (associated with tensor products) of individual path-amplitudes over the separate streams.

By postulate 2, the amplitude of a tangible particle's shadow stream is the sum (integral) of the separate path-amplitudes for particles comprising the stream. This, as indicated above, is what underlies single-particle interference, and it leads to elegant resolutions of single-particle interference paradoxes (although not discussed further in this paper). Notice, too, that the sum is only over path-amplitudes associated with particles in the stream. *To obtain the composite amplitude involving amplitudes coming from different streams (that is, generated by different tangible particles) the separate amplitudes must be multiplied by Postulate 3.*

As stated, the main arguments in this paper focus on one- and two-particle interferometry, where there are only a finite number of alternative paths available to each particle, tangible or shadow. But the general SI system (which may encompass uncountably many paths for a particle traveling between two points) is implemented mathematically by Feynman path integrals, as noted earlier. Thus consider now a single non-relativistic tangible particle, with its shadow stream propagating over an infinitesimal time interval as shown in Fig. 2.1 above. In this case the stream's amplitude, $\psi(x,t)$, is given by a Feynman path-sum (integral). The recursive expression for the (un-normalized) amplitude is, [24]

$$\psi(x,t+\varepsilon) \sim \sum_a \left( \exp[\frac{i}{\hbar} S(x,a)] \right) \psi(a,t). \tag{2.5}$$

Here $S(x,a) = \int_t^{t+\varepsilon} \left( \frac{m}{2}(\frac{d\gamma}{d\tau})^2 - V(\gamma(\tau)) \right) d\tau$ is the action in going from $a$ to $x$ along the path $\gamma$, and $\varepsilon$ is infinitesimal ($V$ is a potential). Also of course, which is implicit in the exponentiated action, we are assuming that shadow particles obey the same dynamics as tangible particles, as mentioned in Sec. I.

In [25, 26] Feynman derived Schrödinger's wave equation from (2.5):

$$i\hbar\frac{\partial\psi}{\partial t} = -\frac{\hbar^2}{2m}\frac{\partial^2}{\partial x^2}\psi + V\psi.$$

As can be seen from Fig. 2.1, a "classical" picture of particle interaction underlies the shadow system. Nevertheless, to emphasize a previous point: Because of Feynman's derivation, the shadow system is a consistent extension of standard quantum mechanics. [27] When we deal with multi-particle systems in Sec. III, this has the important consequence that violation of Bell's inequalities are automatic in the shadow system.

As originally developed, Feynman's path-integral approach seems to require that a particle must simultaneously explore all possible paths in traveling between two points (as in the above quote from Brian Greene; or as Feynman once put it in a different context, the particle "smells all the paths in the neighborhood"). [28] This is certainly a paradox, requiring a particle to move at superluminal speeds. In fact this suggestion of paradox may account for the formulation's being generally overlooked in introductory courses on quantum mechanics, which usually favor the Schrödinger differential-equation approach. But as Jon Ogborn and E. F. Taylor point out,

> Feynman's crucial and deep discovery was that you can base quantum mechanics on the postulate that [the difference between kinetic and potential energy] divided by the quantum of action $h$ is the rate of rotation of the quantum arrow [unit vector in the complex plane]. [29]

In any case, Feynman's path-integral approach has considerable intuitive beauty and pedagogical merit, as E. F. Taylor has shown. [30, 31, 32] Moreover, it quantifies an elegant picture of quantum foundations in Euclidean space, SI: in which the stream is the basic object, and where streams, not particles, are in superposition—a visualizable and intuitively appealing picture.

## III. HOW FEYNMAN'S CLOCKS CORRELATE BERTLMANN'S SOCKS

> Indeed the hidden-variables theories ruled out by Bell's Theorem rest on assumptions that not only can be stated in entirely nontechnical terms but are so compelling that the establishment of their falsity has been called [33], not frivolously, " the most profound discovery of science." [34]

> Anybody who's not bothered by Bell's theorem has to have rocks in his head
> — Arthur Wightman, major contributor to quantum field theory [35]

Most striking is the case of entanglement, which Einstein called "spooky," as it implies that the act of measuring a property of one particle can instantaneously change the state of another particle no matter how far apart the two are … . John

Bell showed that the quantum predictions for entanglement are in conflict with local realism. From that "natural" point of view [local realism] any property we observe is (a) evidence of elements of reality out there and (b) independent of any actions taken at distant locations simultaneously with the measurement. Most physicists view the experimental confirmation of the quantum predictions as evidence for nonlocality. But I think that the concept of reality itself is at stake … . [36]

Truly, something important is at stake, and in this section we will show how the denial of EPNT (BGHZ's "emptiness of paths not taken," from the quote in Sec. I) results in a consistent extension of standard quantum mechanics in position/momentum space that nevertheless possesses local causality (by "local causality," or by what Bell termed "locally explicable," [37] we mean contiguous interaction)—thus overturning one of the supposed consequences of Bell's theorem.

The argument is quite direct and elementary. The only possible issue that could be raised is whether the postulates in Sec. II are consistent with standard quantum mechanics. But to reiterate earlier points from Sec. I: The consistency is immediately clear, almost by definition. The postulates are all just direct translations of the usual properties governing amplitudes in coordinate space (e.g., rules (2.1)–(2.4) in Sec. II, taken from Feynman's text [14]). The basic object in SI is that of a stream generated by a single tangible particle and its accompanying shadow particles (postulate 1). Feynman's path sums (integrals) quantify stream amplitudes (postulate 2), and given two distinct tangible particles, generating two different streams, the amplitude for one stream doing one thing and the other stream doing something else is the product of the separate amplitudes (postulate 3). The predictions of the system only involve such amplitudes, and the outcomes are necessarily consistent with standard quantum mechanics—indeed, as stated, almost by definition. Alternatively, as mentioned in Sec. I, any experimental arrangement in Euclidean configuration space—for example the double-slit setup—is a model of QM + SI. In other words, all true statements of quantum mechanics plus those of SI are satisfiable in the structure, and therefore by a well-known result in mathematical logic the combined theory is consistent. [38]

Now consider BGHZ's standard two-particle interferometer (Rarity-Tapster [39]) shown in Fig. 3.1 below, where a source event (for example, spontaneous down conversion for photons or, say, the decay of a mother particle when other particles are used) creates a pair of tangible daughter-particles. (The argument is quite general for all such arrangements where entangled particles move in coordinate space, and it applies, *mutatis mutandis*, not just to a Rarity-Tapster setup but to any interferometer or double-slit arrangement). Phase shifters $\alpha$ and $\beta$ correspond to the "measurements" carried out independently in the wings of the device.

As is well known (also confirmed by BGHZ), the tangible daughter-particles will take paths *a-a′* or *b-b′*. Of course the position of the source when the daughter particles are emitted will necessarily vary randomly from trial to trial, but we will assume that the angle subtended by, say, *a-b′* is sufficiently small (or equivalently, as it is usually put, the angle subtended by *b′-a′* is sufficiently large [40]) that two-particle interference occurs. In the interferometer there are thus just two choices for the upper tangible particle, either path *a* or path *b′*.

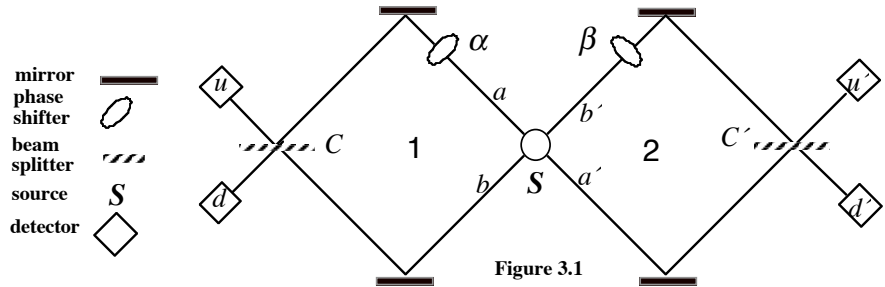

Figure 3.1

Suppose, without loss of generality, that the upper tangible takes path $a$. Therefore the second tangible takes path $a'$. The shadow counterparts then, by postulates 1 and 2, take paths $b'$-$b$. We will let $t$ and $s$ stand for "tangible" and "shadow," with a subscript indicating which path the particle takes. Thus, for example, the tangible particle taking path $a$ is $t_a$ and its shadow counterpart is $s_{b'}$, since the shadow counterpart takes path $b'$ by postulate 1.

The forgoing source event therefore creates four particles and two streams. The particles $t_a$ and $s_{b'}$ are in the same shadow stream, as are $s_b$ and $t_{a'}$. But $t_a$ and $s_b$ are in different streams; so, too, are $s_{b'}$ and $t_{a'}$, by postulates 1 and 2.

Nevertheless, when $t_a$ and $s_b$ meet at a beamsplitter their associated path-amplitudes or path-clocks contain complementary information about each particle's path through the left-half of the interferometer; similarly for $s_{b'}$ and $t_{a'}$ regarding paths on the right-side. Because $t_a$ and $s_b$ come from different streams, when they interfere the information is shared by addition of clock-values, phases—since the amplitude of the composite stream is quantified by multiplication of path-amplitudes (postulate 3). Similarly for $s_{b'}$ and $t_{a'}$. (Here the term "information" is used in a sense similar to the way it is used in Brownian motion, a Markovian process determined by contiguous physical interaction. In such a process the present state's "information" determines the next state—its probability—independently of past states. Contiguous interaction (local causality) at a beamsplitter in SI also determines the probable outcome; but, on the other hand, in SI the rotations of path-clocks between successive intersection points contribute to the interference result at a beamsplitter.)

So far, however, all particle interaction is confined to each side separately. But notice that the information carried by $s_b$ and $t_{a'}$ (stored in their respective path-amplitudes or clock values) is the same, since they take congruent paths, [41] although the particles are confined to opposite sides. Nevertheless, on each side of the interferometer, when the separate particles encounter a beamsplitter, where they go next is determined by these clock values. But here is the key property of the shadow system that leads to correlation: The information stored in the associated clock values or path-amplitudes of $s_b$ and $t_{a'}$ is the same. It is this information that determines the interference effects by means of contiguous interaction.

Now for the formal calculation, which simply implements the forgoing comments mathematically.

Let $\langle u,u'|\psi_0 \rangle$ be the amplitude that the tangible-daughter pair arrives at detectors $u$ and $u'$, and let $\langle u,d'|\psi_0 \rangle$ be the corresponding amplitude for arrival at $u$ and $d'$ ($\psi_0$ is the initial state of the system). In all, there are four possible amplitudes for arrival at the detectors. The other two are $\langle d,u'|\psi_0\rangle$, and $\langle d,d'|\psi_0\rangle$). Without loss of generality, we will content ourselves with just calculating $\langle u,u'|\psi_0\rangle$, and $\langle u,d'|\psi_0\rangle$.

Working in the tensor-product space, we see that the amplitude $\langle u,u'|\psi_0\rangle$ is the sum of two amplitudes, $\langle u|a\rangle\langle u'|a'\rangle$, and $\langle u|b\rangle\langle u'|b'\rangle$. This calculation is carried out in the standard way, and the result makes sense from our postulates and from our perspective of physical realism because, as BGHZ have established, the pair of daughter tangible-particles travel either the paths $a$-$a'$ or the paths $b$-$b'$. Moreover, each of the summands is a product (tangible and shadow particles coming from different streams, postulate 3)—which is of course how amplitudes are calculated in the tensor-product space. Thus amplitude for $u$ - $u'$ : $\langle u,u'|\psi_0\rangle = \langle u|a\rangle\langle u'|a'\rangle + \langle u|b\rangle\langle u'|b'\rangle$

$$= \left(\frac{1}{\sqrt{2}}e^{i\alpha}i\right)\left(\frac{1}{\sqrt{2}}\right) + \left(\frac{1}{\sqrt{2}}\right)\left(\frac{e^{i\beta}}{\sqrt{2}}i\right) = \frac{i}{2}(e^{i\alpha} + e^{i\beta}) \quad (3.1)$$

amplitude for $u$-$d'$: $\langle u,d'|\psi_0\rangle = \langle u|a\rangle\langle d'|a'\rangle + \langle u|b\rangle\langle d'|b'\rangle$

$$= \left(\frac{1}{\sqrt{2}}e^{i\alpha}i\right)\left(\frac{1}{\sqrt{2}}i\right) + \left(\frac{1}{\sqrt{2}}\right)\left(\frac{e^{i\beta}}{\sqrt{2}}\right) = \frac{1}{2}(-e^{i\alpha} + e^{i\beta}) \quad (3.2)$$

Equations (3.1) and (3.2) yield the standard probability outcomes:

$$P_r(u,u') = \langle u,u'|\psi_0\rangle^* \langle u,u'|\psi_0\rangle$$
$$= \frac{1}{4}(e^{-i\alpha} + e^{-i\beta})(e^{i\alpha} + e^{i\beta}) = \frac{1}{4}\left(2 + e^{-i(\beta-\alpha)} + e^{i(\beta-\alpha)}\right) = \frac{1}{2}\cos^2\frac{\beta-\alpha}{2} \quad (3.3)$$

$$P_r(u,d') = \langle u,d'|\psi_0\rangle^* \langle u,d'|\psi_0\rangle$$
$$= \frac{1}{4}(-e^{-i\alpha} + e^{-i\beta})(-e^{i\alpha} + e^{i\beta}) = \frac{1}{4}\left(2 - e^{-i(\beta-\alpha)} - e^{i(\beta-\alpha)}\right) = \frac{1}{2}\sin^2\frac{\beta-\alpha}{2} \quad (3.4)$$

On the face of it, the equations leading to (3.1) and (3.2) have the appearance of non-local causality: For how can a particle moving along paths within one side of the interferometer coordinate its travels with a particle moving on the other side? Indeed, we have just seen that the inequalities are violated in our computation. So it seems to follow from Bell's theorem that local causality must fail. As the usual story goes (incorrectly), there must be instantaneous communication across arbitrarily separated spatial regions. Indeed, as Bell put it (quoted in Sec I) the correlations must be "… *locally inexplicable*. They cannot be explained, that is to say, without action at a distance."

But wait! This is where shadow particles make all the difference: As pointed out above, the shadow particle $s_b$ carries the same information as the tangible particle $t_{a'}$. The accumulated information is stored in each particle's path-amplitude or path-clock (phase) as the particle travels through the interferometer. The information is the same for the simple reason that the particles take congruent paths (though as pointed out in [36], path-amplitude equivalence suffices here). Thus,

$$\langle u|b\rangle = \langle u´|a´\rangle \text{ and } \langle d|b\rangle = \langle d´|a´\rangle. \tag{3.5}$$

Therefore making the substitutions from (3.5) into (3.11) and (3.2) we have,

$$\langle u|a\rangle\langle u´|a´\rangle + \langle u|b\rangle\langle u´|b´\rangle = \langle u|a\rangle\langle u|b\rangle + \langle u´|a´\rangle\langle u´|b´\rangle \tag{3.6}$$

$$\langle u|a\rangle\langle d´|a´\rangle + \langle u|b\rangle\langle d´|b´\rangle = \langle u|a\rangle\langle d|b\rangle + \langle u´|a´\rangle\langle d´|b´\rangle \tag{3.7}$$

As we see, in the second half of each of the equations (3.6) and (3.7) the information exchanged is purely local in nature (via contiguous interference effects) when the tangible and shadow particles interact at the beamsplitters. Thus, although the interaction of tangible and shadow particles at a beamsplitter has a random component (postulate 2), it determines—in each separate side of the device—which detector is reached by the tangible particle, and that in turn determines which detectors are reached by the pair of tangible particles traveling on opposite sides of the interferometer. Each separate particle proceeds according to purely local influences, its ultimate destination determined by the shared information between the tangible and shadow particles. There is no action at a distance, although the end result is as though there were. **QED**

Central to the forgoing argument is that when there is more than one possible path between two points, the tangible particle *randomly* takes one of the paths and shadow particles take the others. This property of the system is a perfect match for Feynman's path-integral formulation of quantum mechanics. Furthermore, it has the consequence that in the shadow system the outcome for entangled particles is inherently random, not predetermined when the source event occurs. As the shadow and tangible particles travel their separate paths through the interferometer, their "clocks" or phases are initially correlated but set randomly and go on to coordinate precisely the interference at a beamsplitter, but where the particles ultimately go next (to which detectors) is a random outcome of interaction with the beamsplitter. Nevertheless, the trajectories of tangible and shadow particles in SI have pre-existing values at each instant along a path. That is, there is local realism; and, as we have seen, local causality operates throughout from source event to arrival at a detector.

Because the shadow system is intrinsically random, the original EPR program (Sec. I) is not completely vindicated (though randomness was not an important issue for Einstein) [42]. Nevertheless, pronouncements such as this recent one by Alain Aspect (who is famous for having carried out the first experimental realization of the EPR-Bohm gedanken experiment) are no longer tenable: [43]

> … after Bell's discovery that local realism entailed a limit on the correlations—a limit he expressed in his celebrated inequalities—a series of ever more ideal experiments has led us to abandon the concept [local realism]. It is then natural to raise the question of whether one should drop locality—which equates to the impossibility of any influence traveling faster than light—or rather drop the notion of physical reality.

Indeed, the forgoing derivation presents a striking dichotomy for the choices of interpretations in quantum mechanics. The BGHZ argument, although independent of ours, can be viewed as a carefully reasoned and valid proof of the implication (in Euclidean configuration space),

$$\text{EPNT} \Rightarrow \text{non-local causality.}$$

But the above argument shows, both conceptually and mathematically,

$$\text{not-EPNT (in the form of SI)} \Rightarrow \text{local causality.}$$

This leaves us with only two viable choices for quantum foundations (at least with respect to spinless particles; but see the remark about spin at the end of Sec. I): systems harboring instantaneous action across spatially separated regions or those containing hypothetical "somethings"—or, if you will, "spooky action-at-a-distance" versus "spooky shadow particles." But as John Wheeler and Martin Gardner point out [44] for action-at-a-distance the particles must "remain connected, even though light years apart, by a nonlocal sub-quantum level that no one understands … " As the old saying goes: You pays your money, and you takes your choice—but for one choice the cost seems rather high.

[1] H. J. Bernstein, D. M. Greenberger, M. A. Horne, A. Zeilinger, "Bell theorem without inequalities for two spinless particles," Physical Review A, Vol. 47, Number 1, January, 1993

[2] J. G. Rarity, P. R. Tapster, "Experimental Violation of Bell's Inequality Based on Phase and Momentum," Physical Review Letters, Vol. 65, Number 21, 21 May 1990

[3] A. Einstein, B. Podolsky, N. Rosen, "Can quantum mechanical description of reality be considered complete?" Phys. Rev. 47, 777 (1935).

[4] D. Deutsch, *The Fabric of Reality*, Penguin Press, 1998

[5] J. S. Bell, "Bertlmann's socks and the nature of reality," in *Speakable and Unspeakable in Quantum Mechanics*, 2nd ed., Cambridge University Press, 2004

[6] J. S. Bell, "On the Einstein-Podolosky-Rosen Paradox," Physics I , 195-200 (1964) (reprinted in Ref. 5)

[7] R. P. Feynman, QED: *The Strange Theory of Light and Matter* (Princeton University Press, Princeton, NJ, 1986).

[8] D. M. Greenberger, M. A. Horne, A. Zeilinger, "Multparticle Interferometry and the Superposition Principle," Physics Today, August 1993

[9] R.P. Feynman, A. R. Hibbs, *Quantum Mechanics and Path Integrals*, McGraw-Hill, New York, 1965

[10] Herbert B. Enderton , "A Mathematical Introduction to Logic": Second Edition, 2001, Harcort Academic Press, Burlington MA.

[11] W. Tittel, J. Brendel, B. Gisin, T. Herzog, H. Zbinden and N. Gisin, "Experimental demonstration of quantum correlations over more than 10 km," Physical Review A **57**(5): 3229 - 3232, 1 May 1998

[12] N. Gisin, V. Scarani, A. Stefanov, A. Suarez, W. Tittel and H. Zbinden, "Quantum Correlations With Moving Observers," Optics & Photonics News, 51, December 2002

[13] L. Gilder *The Age of Entanglement : When Quantum Physics was Reborn*. New York : Alfred A. Knopf, 2008.

[14] Wall Street Journal, May 5, 2009

[15] L. Vaidman, "Many Worlds Interpretation of Quantum Mechanics," Stanford Encyclopedia of Philosophy, 2002

[16] L. Schulman, *A Path Integral for Spin*, Phys. Rev. 176, 1558-1569 (1968)

[17] A. Altland, B. Simons, *Condensed Matter Field Theory, Cambridge University Press*, 2006

[18] H. Kleinert, Path Integrals in Quantum Mechanics, Statistics, Polymer Physics, and Financial Markets, 2009, World Scientific 5th edition

[19] B. R. Greene, *The Elegant Universe*, Vintage, 1999

[20] Ref. 9, ch. 2

[21] R. P. Feynman, R. B. Leighton, and M. Sands, *The Feynman Lectures on Physics* (Addison-Wesley, Reading, MA, 1964), Vol. III, Ch 3

[22] R.P. Feynman, "Space-Time Approach to Non-Relativistic Quantum Mechanics," Rev. Mod. Phys. 20, 367-387 (1948) reprinted in *Feynman's Thesis, A New Approach to Quantum Mechanics*, edited by Laurie M. Brown, World Scientific Publishing, 2005.

[23] H. J. Keisler, *Elementary Calculus An Infinitesimal Approach*, Prindle, Weber and Schmidt, 1976

[24] Ref. 22

[25] Ref. 22

[26] Feynman's Nobel prize acceptance speech contains a fascinating personal account of how he arrived at his result.

[27] BGHZ elaborated on EPNT in a supplementary paper, "Further evidence for the EPNT Assumption," *NASA. Goddard Space Flight Center, Third International Workshop on Squeezed States and Uncertainty Relations*, p 535-543 (SEE N95-13896 03-74). But their argument in favor of EPNT is not consistent with Feynman's construct. They assume that the amplitude for a path not taken by the tangible particle is zero; but, following Feynman, we attach an amplitude to all *possible* paths (see equation (2.5).

[28] Ref. 21, Vol. II, Ch 19

[29] J. Ogborn, E. F. Taylor, Physics Education, January, 2005.

[30] E. F. Taylor, S. Vokos, J. M. O'Mearac, N. S. Thornber, "Teaching Feynman's Sum-Over-Paths Quantum Theory," Computers In Physics, Vol. 12, No. 2, Mar/Apr 1998

[31] E. F. Taylor, "A call to action," Am. J. Phys. **71** 5, May 2003

[32] E. F. Taylor, "Rescuing quantum physics from atomic physics," www.eftaylor.com

[33] H. Stapp, Nuovo Cimento, 40B, 191 (1977)

[34] N. D. Mermin, "Hidden variables and the two theorems of John Bell," Reviews of Modern Physics, Vol. 65, No. 3, July 1993

[35] Quoted by N. D. Mermin, *Is the Moon there when nobody looks? Reality and the quantum theory*, Physics Today, April, 1985

[36] A. Zeilinger, "The message of the quantum," Nature, Vol 438|8, December 2005

[37] Ref. 5

[38] Ref. 10

[39] Ref. 2

[40] D. M. Greenberger, M. A. Horne, A. Zeilinger, "Multparticle Interferometry and the Superposition Principle," Physics Today, August 1993

[41] Of course congruence is not a necessary condition, only sufficient. As a practical matter, in designing the interferometer, the experimenters (operating under the principle of path-indistinguishability) usually construct the device so that the paths have the same length (are congruent). As mentioned in Sec. I, however, the necessary condition is that certain path-amplitudes (as in equation (3.5)) must be the same—what we call, equivalent—in order for two-particle interference to occur.

[42] Ref. 35

[43] A. Aspect, "To be or not be local," NATURE Vol 446|19, April 2007

[44] M. Gardner, J. A. Wheeler, *Quantum Theory and Quack Theory*, New York Review of Books, Vol. 26, Number 8, May 17, 1979